# Ultrafast and reliable domain-wall and skyrmion logic in a chirally coupled ferrimagnet

Yifei Ma[1,2†], Dihua Wu[3†], Fengbo Yan[1], Xiaoxiao Fang[2], Peixin Qin[4], Leran Wang[2], Li Liu[4], Laichuan Shen[5], Zhiqi Liu[4], Wenyun Yang[2], Jie Zhang[6], Yan Zhou[5], Feng Luo[1*], Jinbo Yang[2,7*], Hyunsoo Yang[8], Kaiming Cai[3*], Shuai Ning[1*], Zhaochu Luo[2,7*]

[1]Tianjin key lab for rare earth materials and applications, Center for rare earth and inorganic functional materials, School of materials science and engineering, Nankai University, 300350 Tianjin, China.
[2]State key laboratory of artificial microstructure and mesoscopic physics, School of physics, Peking University, 100871 Beijing, China.
[3]School of physics, School of integrated circuits, Huazhong University of Science and Technology, 430074 Wuhan, China.
[4]School of materials science and engineering, Beihang University, 100083 Beijing, China.
[5]School of science and engineering, The Chinese University of Hong Kong, 518172 Shenzhen, China.
[6]School of computer science, Peking University, 100871 Beijing, China.
[7]Beijing key laboratory for magnetoelectric materials and devices, 100871 Beijing, China.
[8]Department of electrical and computer engineering, National University of Singapore, 117576 Singapore.
†These authors contributed equally: Yifei Ma, Dihua Wu.
*Corresponding authors: zhaochu.luo@pku.edu.cn; kmcai@hust.edu.cn; sning@nankai.edu.cn; feng.luo@nankai.edu.cn; jbyang@pku.edu.cn.

**SUMMARY：**

Unlocking the spin degree of freedom in addition to the electron's charge, spin-based logic offers an in-memory computing architecture beyond-CMOS technology. Here, we encode information into chiral spin textures (e.g., chiral domain-wall and skyrmion) and achieve an ultrafast and reliable all-electrical logic by exploiting the Dzyaloshinskii-Moriya interaction-induced chiral coupling. Taking advantage of fast spin dynamics in antiferromagnetically coupled systems, we achieved a fast domain-wall motion passing through the logic gate, exceeding 1 kilometre per second, yielding an operation time of 50 picoseconds for a 50 nanometres-long logic gate. Furthermore, we present a fast logic operation with skyrmion bubbles in a racetrack that exhibits a topologically protected computation scheme. Our work demonstrates a viable approach for advanced microchips with high operation frequency and ultralow power consumption, paving the way for next-generation computing technologies.

## INTRODUCTION

Modern computers, following the Von Neumann architecture, have a hierarchical structure spanning from the central processing unit (CPU) to multi-level storage systems, with an inverse relationship between memory capacity and operation speed (Figure 1A). While this computer architecture has achieved great success, it faces several bottlenecks as Moore's law approaches its physical limits. On one hand, due to the low carrier mobility and high energy consumption in transistors, the clocking frequency of CPU has stuck at around several GHz for decades. On the other hand, silicon-based CPU and low-level memories are volatile and data need to be frequently transferred with slow non-volatile memory unit across a large time scale, leading to the notorious von Neumann bottleneck. Owing to the built-in non-volatility, spin-based logic devices provide a beyond-CMOS platform for scalable in-memory computing that outperforms conventional von Neumann computer in solving emergent tasks associated with artificial intelligence (AI).[1-7] Among various spin-based logic devices, spin-texture racetrack logic, which encodes the information in nanosized spin textures such as chiral domain wall (DW) and magnetic skyrmion—a swirling chiral spin texture with topologically protected stability—attracts intensive attention due to its advantages in easy cascading for forming large-scale logic circuits[3,6] and seamless integration with high-density racetrack memory (Figure 1B).[8,9] In order to obtain efficient and fast logic operation, current-driven spin-texture motion with a high velocity in the logic device is highly demanded. So far, limited by the inherent spin dynamics of ferromagnets, the velocity for a DW passing through the logic device is around a hundred meter per second, giving an operation frequency of a few GHz.[6] Antiparallel exchange-coupled systems such as synthetic antiferromagnets,[10] antiferromagnets[11], ferrimagnets[12-16] and synthetic ferrimagnets[17] exhibit faster spin dynamics, offering promising candidate materials for high-speed spin-based logic.

However, a mechanism for realizing racetrack logic in antiparallel exchange-coupled systems remains elusive, and fast electrical manipulation of topological spin textures has not yet been experimentally explored in the regime of logic operation. In this work, by exploiting interfacial Dzyaloshinskii-Moriya interaction (DMI)-induced lateral magnetic coupling, we demonstrate an all-electrical DW and skyrmion logic in the ferrimagnetic CoGd alloy. We adjust the atomic composition of the ferrimagnet close to the angular momentum compensation to obtain a high DW velocity in the DW NOT gate. Moreover, we employ this concept into magnetic skyrmion bubbles and demonstrate a fast skyrmion NOT gate with preserved topological winding number.

## RESULTS

### Logic device construction from chirally coupled ferrimagnets

In the ferrimagnetic CoGd alloy, wherein the Co and Gd spin sublattices exhibit strongly antiparallel exchange coupling, its net magnetization ($M_{net}$) and angular momentum ($S_{net}$) are given by:[12-15] $M_{net} = M_{Co} - M_{Gd}$, $S_{net} = S_{Co} - S_{Gd}$, where $M_{Co(Gd)}$ and $S_{Co(Gd)}$ represent the magnetization and angular momentum of the Co (Gd) sublattice, respectively. $M_{net}$ and $S_{net}$ depend on the temperature and atomic composition, which can reach their fully compensated state from the control of temperatures or compositions. Notably, in the vicinity of the angular momentum compensation, high field- and current-driven DW velocities have been reported as a consequence of fast antiferromagnetic spin dynamics.[12-15] Yet the implementation of such fast dynamics used for logic operations has not been realized. To investigate the compensation state in ferrimagnets, magnetic multilayers of Ta (1 nm)/Pt (5 nm)/Co$_{1-x}$Gd$_x$ (7 nm)/Ta (2 nm) ($x$ =20~30 at%) are grown by magnetron sputtering (details seen in Methods) (Figure 1C). The Pt layer can provide spin-orbit torques (SOTs)[18] and large interfacial DMI.[13,15] By measuring the saturation magnetization and coercivity, we determine the compositions for the magnetization and angular momentum compensation at room temperature are approximately 23.5% and 22.1% of Gd, respectively (Supplementary Note 1).

In order to construct spin-texture logic devices, focused ion beam (FIB) technique is employed to precisely modulate the magnetic properties of CoGd films (Figure 1C) (Supplementary Note 2). The as-grown ferrimagnetic film exhibits an excellent perpendicular magnetic anisotropy (PMA) (Figure S3). By locally irradiating ferrimagnets with various Ga ion energies, the PMA gradually declines (Figure 1D), which could be attributed to the elemental mixing at the Pt/CoGd interface.[19,20] As shown in Figures 1E and F, each interface in the as-grown multilayer is sharp, and Co and Gd atoms are uniformly mixed. By contrast, after the irradiation, the interfaces become blurred and a noticeable interdiffusion occurs at interfaces (Figure 1G). Since the PMA partially originates from the spin-orbit coupling at the Pt/CoGd interface and the atomic microstructure,[21,22] the degradation of the interface will diminish the PMA (Figure S4).

Utilizing the high-resolution (<10 nm) FIB technique, we pattern nano-sized regions with the in-plane (IP) magnetic anisotropy in an out-of-plane (OOP) magnetized CoGd films and demonstrate the DMI-induced chiral coupling in the ferrimagnetic system. As illustrated in Figure 2A, due to the presence of alternating OOP-IP-OOP magnetic anisotropies, it prefers to form noncollinear spin textures. As a consequence of the energetical competition between the exchange interaction, magnetic anisotropy as well as DMI, the noncollinear spin texture within the OOP-IP-OOP structure follows the chiral Néel-type configuration, such as ⊙←⊗ and ⊗→⊙ (⊙ or ⊗ indicates the up or down direction of the unit Néel vector defined as $\boldsymbol{n}$ = ($\boldsymbol{m}_{Co}$ - $\boldsymbol{m}_{Gd}$)/2 in CoGd), leading to a preferable antiparallel alignment between neighbouring OOP regions.[23] We quantify the strength of the chiral coupling by studying the hysteresis behaviour (Supplementary Note 3). The coupling strength is estimated to be (1.14±0.22)×10$^{-4}$ pJ, corresponding to a DMI constant of $D$ = 0.22±0.04 mJ/m$^2$ which aligns with previous estimates in Pt/CoGd systems.[13,15]

**High-performance DW logic operation**

The lateral magnetic coupling can be utilized to design many functional devices such as DW logic,[6] magnetic memory,[24,25] spin diode[26] and Ising machine.[27] To demonstrate fast current-driven DW inverter, equivalent to a NOT gate, we patterned the ferrimagnetic racetrack with a 50 nm-wide IP region in our experiments. As illustrated in Figure 2B, on application of current pulses in Pt layer, the SOTs can drive the DW motion along the racetrack in the absence of magnetic fields. When encountering the IP region, the incident DW annihilates and a reversed domain nucleates on the other side of the IP region, making the spin alignment around the OOP-IP-OOP preserve the same chirality. In the experiment, the DW motion is monitored with the polar magneto-optic Kerr effect (MOKE) microscopy with visible light wavelength that detects the direction of Co magnetic moments (Figure 2C).[28,29] As shown in Figure 2D, after passing through the IP region, the polarity of the DW is inverted into ⊙|⊗. An analogous inversion process occurs for an incident ⊙|⊗ DW as well as the reversed currents (Figures 2D and S6). This process transforms the ⊙|⊗ (⊗|⊙) DW into a ⊗|⊙ (⊙|⊗) DW, verifying the functionality of DW inversion.

In DW-based devices, the key performances greatly depend on the DW motion speed. We first estimated the DW velocity from the current-driven DW displacement over a certain pulse duration. As shown in Figure 2E, the DW velocity increases rapidly and reaches a maximum of 1.10±0.02 km/s (for a current density $J$ =1.10×10$^{12}$ A/m$^2$). The non-monotonic relation can be attributed to the temperature-dependent magnetic properties from self-heating.[15] We choose the atomic composition, where the angular momentum compensation temperature ($T_{AMC}$) is slightly higher than room temperature, to obtain a fast DW motion, which could be further enhanced from material engineering.[15,16] In particular, when Gd atomic composition is ≥23%, the maximum DW velocity is higher than 1 km/s over a wide temperature range from 290 to 370 K in our experiments, indicating a good stability for electronic devices. We further estimate the effective velocity ($v_{INV}$) for the DW passing through an inverter as a function of the current density (details seen in Methods of Supporting Information) (Figure 2E). Notably, $v_{INV}$ can reach a maximum value of 1.01±0.08 km/s, which is almost one order of magnitude higher than that in a ferromagnetic system of Pt/Co.[6] Besides, the fast logic operation retains no degradation after 10$^9$ operation cycles, verifying a good endurance. Note that the onset current density in the ferrimagnetic DW inverter is higher than that in a uniform racetrack, indicating an additional energy barrier associated with the OOP-IP-OOP structure. This can locally engineer the energy landscape for DW motion, offering a scalable scheme to overcome the trade-off between energy efficiency and thermal stability in DW-based racetrack devices.[30]

**Fast DW inversion and device downscaling**

To incorporate with CMOS technology, magnetic tunnel junctions (MTJs) can be implemented on top of the racetrack for reading the input/output, while transistors for current injection in the racetrack to execute write operation,[31] as illustrated in Figure 3A. To understand the experimental observations and working principles, we performed the numerical simulations of an antiferromagnetically coupled system with two sublattices (details seen in Methods). A peak velocity is obtained with $x_{Gd}$ ~22% at the compensated ferrimagnet (Figure 3B). Specifically, a DW speed of ~2.3 km/s is obtained for a current density of 2.0×10$^{12}$ A/m$^2$. To elucidate the mechanism of DW inversion, we investigate the DW dynamics in the ferrimagnet inverter (Figure 3C). When the incident DW (referred to as the input DW) reaches the IP region, the region between two domains exhibits opposite spin alignments, corresponding to a strong repulsion and high-energy states. Consequently, the input DW decelerates and finally gets pinned in IP region, while a DW with opposite polarity (referred to as the output DW) starts to move towards the end of the right side. We extract the time-dependent locations of the input and output DWs (Figure 3D). At low current densities, the input DW cannot penetrate the IP region. Above a certain current density, the output DW is generated with nearly zero delay time. Compared with the ferromagnet system of Pt/Co, the ferrimagnetic inverter shows a much faster $v_{INV}$ and higher onset current density, bringing to a more reliable performance (Figure 3E).

To study the device downscaling, we first define the geometry of the device and operation time. An inverter consists of OOP-IP-OOP regions as a total length ($L_{INV}$) and the operation time as the period of DW transferring through the gate. In our experiment, with $L_{INV}$ = 50 nm, an estimated operation time is ~50 ps for $v_{INV}$ ~1.0 km/s. Miniaturization of the devices with reducing racetrack length and increasing the DW velocity can further improve the operation speed. Besides, the ferrimagnet shows a narrow DW width and has an ultrafast relativistic speed limitation owing to a high maximum magnon group velocity.[16] From the simulation, we show that the downscaled ferrimagnetic inverter can retain good functionality even at a small dimension of $L_{INV}$ =10 nm, giving an extremely high operation frequency of ~100 GHz, potentially superior to the advanced CMOS gates in the operation frequency and energy efficiency (Supplementary Note 8). Similarly, we have also demonstrated reconfigurable NAND and NOR logics which is functionally complete and able to construct arbitrary logic functions (Supplementary Note 9). Unlike the CMOS gate suffering from the parasitic capacitance/inductance-induced trade-off between the operation speed and energy efficiency, our DW logic can simultaneously achieve sub-THz-level operation frequency and ultralow power consumption. Moreover, due to the long retention of the data stored in DWs, nearly zero static power consumption could be expected in our DW logic without frequently refreshing. From the aspect of operation speed and energy consumption, our DW logic device is highly competitive among the existing in-memory computing schemes (Figure S15), potentially addressing the "power-wall" issue and enabling higher clock frequency and integration density in future microchips.

**Reliable current-driven skyrmion NOT logic**

As a one-dimensional spin texture, the DW motion is prone to the edge defects, resulting in potential information loss when multiple DWs become pinned or annihilated at these defects (Figures S21A and D). To mitigate this issue, we further explore the reliability of logic operation with magnetic skyrmions for applications, such as in-memory computing. Magnetic skyrmions are particle-like swirling spin textures with topological protection, and compared to DWs, possess more freedom to move on a two-dimensional surface (Figures S21B-F), providing a promising candidate for high-density storage and emergent computing.[32,33] In spite of several theoretical attempts in skyrmion-based logic devices,[34-37] the experimental demonstration remains elusive. We generate the magnetic bubbles in the ferrimagnet via the interplay of SOTs, magnetic fields and chiral coupling (Figure S17). On application of current pulses, the magnetic bubble can move steadily along the direction of current, verifying the Néel-type spin alignment of a magnetic skyrmion.[32,33] The skyrmions achieve a maximum velocity of 0.83±0.02 km/s at a current density of $J$ = 1.05×10$^{12}$ A/m$^2$ (Figure 4A). Similar to the DW logic gate, we provide an approximate estimation of the operation time for a skyrmion bubble NOT logic is approximated as $t_{SK\_NOT}$ ≈ 280 ps (see more details in Supplementary Note 13). Compared to DWs in the same ferrimagnetic films, the skyrmion bubbles have a slightly lower velocity due to possible deformation at high currents, while maintaining a comparable mobility (defined as the ratio of velocity to current density). The maximum mobilities of DW and skyrmion to be 8.8×10$^{-10}$ and 8.5×10$^{-10}$ m$^3$/(A·s), which is similar to those was observed previously with ferrimagnetic systems,[38] where the mobility is 6.1 and 9.2×10$^{-10}$ m$^3$/(A·s). This mobility is higher than those in the ferromagnetic systems, such as CoFeB [39,40] and Co/Ni/Co,[41] where the mobility is (0.2-1) ×10$^{-10}$ m$^3$/(A·s). Given its topological nature, the skyrmion bubble will be deflected perpendicular to its motion direction by the Magnus force, referred as skyrmion Hall effect. The skyrmion annihilation generally occurs when skyrmion bubbles are pushed to the edge of the racetrack.[42,43] Thanks to the small skyrmion Hall angle, we achieved a precise control of skyrmion bubble motions along the racetrack, minimizing the risk of annihilation and ensuring the stability and accuracy of data flow, which is crucial for practical applications to avoid unexpected errors in data processing (Figures 4B-D and Figure S19).

We then demonstrate the logic operation with skyrmion bubbles (Figures 5A and B). As demonstrated in our previous work,[44] the DW inversion is most effective for narrow V-shaped IP regions with a fan angle, due to the higher energy gain associated with the DMI and lower energy cost of creating a DW in narrow constrictions. As shown in Figure 5C, it creates a gap between two skyrmion bubbles when a skyrmion bubble transmits through the asymmetric geometry of horn-shaped IP region. An analogous inversion process results in the creation of a skyrmion bubble when two skyrmion bubbles separated with a gap sequentially transmit through the IP region (Figure 5D). The relationship between the input and output corresponds to the required operation for a NOT gate (Figure 5B). We also verify the functionality of the NOT gate with small skyrmions by micromagnetic simulation. The small skyrmion with the diameter of ~15 nm can transmit through the NOT gate at a high speed, demonstrating the suitability for fast logic operations (Figures 5C and S20). Moreover, as illustrated in the micromagnetic simulations, the skyrmion bubble first contacts the tip of the horn-shaped IP region, and then gradually completes the inversion process. This horn shape can improve the reliability of skyrmion NOT operation. Notably, the small skyrmion Hall effect in the compensated ferrimagnet allowing for a straight path of skyrmion bubbles' movement towards the logic gate, is critical to realize robust logic operations. In contrast, the skyrmion will be deflected away from the gate structure in the ferromagnetic or uncompensated ferrimagnetic device, prohibiting the execution of skyrmion logic operation (Figure S20).

More interestingly, we find that the topology of the spin structure remains unchanged during the skyrmion logic operation. The topology of a swirling spin structure can be characterized by the topological winding number: $N_w$=- $\frac{1}{4\pi}\int \boldsymbol{n}\cdot(\partial_x\boldsymbol{n}\times\partial_y\boldsymbol{n})dxdy$, where $\boldsymbol{n}$ is the unit Néel vector. By calculating the topological winding number of spin structures obtained in the simulation, we track the evolution of the spin topology during the NOT operation with the logical input of "1" and "0" (Figures 5E and F). Remarkably, the topological winding number is preserved, with values of $N_w \approx$ -1.9 for logic input "1" and $N_w \approx$ -1.4 for inputs "0". Note that in order to keep the topology of the spin structure unchanged during the logic operation, the definition of encoding digital information into the skyrmion bubbles is analogous to that used in DW logic devices (Figure S23). One major advantage of using skyrmion as information carrier is the presence of topological protection. The non-trivial topology invariant during the logic operation implies the extra stability and protection against perturbations from topological nature of magnetic skyrmion. Therefore, albeit the micrometre-sized skyrmion bubbles in the experiment can be downscaled from material optimization, our principle for manipulating skyrmion offers a building block to design fast and reliable computing devices (Supplementary Note 12).[45-48]

**DISCUSSION**

Taking advantage of fast DW velocity and small skyrmion Hall angle of the compensated ferrimagnet, we demonstrate ultrafast and reliable DW and skyrmion logic gates with preserved topological winding numbers. Since both the input and output in our logic devices are encoded into the same type of spin textures, it is straightforward to cascade elementary gates by connecting the input/output racetracks to execute complex tasks. In particular, the fast and efficient in-memory computing capability from our DW and skyrmion-based logic devices can reduce the overhead associated with data transfer, thereby may enabling a substantial acceleration— hundreds of times faster—of compute- and data-intensive applications such as large language models.[49] Additionally, thanks to the built-in non-volatility, our logic devices can persist metadata seamlessly throughout the AI training period, effectively eliminating the troublesome checkpointing process at each training epoch. Therefore, our work provides a route to design extremely fast and efficient in-memory computing microchips for emerging AI tasks with the potential to be directly deployed on the client side.

**METHODS**

**Sample preparation and magnetic properties characterization**

Ferrimagnetic multilayer of Ta (1 nm)/Pt (5 nm)/Co$_{1-x}$Gd$_x$ (7 nm)/Ta (2 nm) ($x$ =20~30 at%) were deposited on thermally oxidized silicon substrates using magnetron sputtering system. The base pressure of the main chamber was better than 5×10$^{-8}$ Torr, and the Ar working pressure was set at 3 mTorr. The ferrimagnetic CoGd alloy layer was deposited by co-sputtering Co and Gd targets. To adjust the atomic composition in CoGd films, the sputtering power of the Gd target varied from 15 to 30 W, while the sputtering power of the Co target was fixed at 75 W. Deposition rates for Co and Gd targets under different sputtering power were calibrated using X-ray reflectivity measurements of Co and Gd control films, and the composition ratio of Co and Gd was calculated according to their density by using the equation of $N_{Co}/N_{Gd} = 3.01 v_{Co}/v_{Gd}$, where $N_{Co(Gd)}$ and $v_{Co(Gd)}$ represent the number of Co (Gd) atoms and the Co (Gd) deposition rate, respectively. The magnetization hysteresis loops were measured using vibrating sample magnetometer (VSM) at room temperature. The Hall bar, domain wall and skyrmion devices were patterned using photolithography and Ar ion milling. The contact electrodes of Ti (10 nm)/Au (50 nm) were fabricated by e-beam evaporation with a lift-off process.

**Process of Ga focus ion beam irradiation**

The Ga ion irradiation process was performed by using the focused ion beam scanning electron microscopy system (FEI Helios 5CX), with different acceleration voltages (5~30 keV) and doses (17~114 μC/cm$^2$). The beam current was about 24 pA. The spot size of the ion beam was about 10 nm, determining the spatial resolution of the irradiation process. To irradiate a large area or a specific pattern using the ion beam, GDSII files were imported into the FIB-NanoBuilder software, and the scanning electron microscope (SEM) system was then used to precisely locate the treated area. Finally, the FIB-NanoBuilder software was executed to achieve the patterned raster-scanned across the sample. The dose $Q$ was given by $Q = \frac{I_B t_D}{LS \times SS}$, where $I_B$ is the beam current, $t_D$ is the dwell time at each position, LS and SS are the line and spot spacing, respectively. The irradiation dose was varied by controlling the dwell time for different devices.

**Microstructure and composition analysis**

Cross-sectional transmission electron microscope (TEM) samples were prepared by focused ion beam system. Before fabricating the TEM samples, amorphous Pt layers were deposited to protect magnetic multilayer from Ga ion implantation. Atomic-resolution cross-sectional TEM imaging and EDS elemental mapping were performed on an aberration-corrected scanning transmission electron microscopy (STEM, FEI Titan cubed Themis G2 300) operating at 200 kV.

**X-ray magnetic circular dichroism measurement**

X-ray absorption spectroscopy (XAS) was conducted in total electron detection mode at the Beamline BL08U1A of Shanghai synchrotron radiation facility (SSRF). The XMCD spectra were recorded with left- and right-hand circularly polarized X-rays with incident light vertical to the film. All X-ray absorption spectroscopy data were taken with normal X-ray incidence and fixed circular polarization. For each measurement, two XAS spectra per helicity were recorded. The XMCD spectrum was obtained from their difference. The Gd XAS background was corrected by linear interpolation of the off-peak intervals $E$ <1,170 eV, 1,194 eV < $E$ <1,206 eV and $E$ >1,230 eV, and the Co XAS background was corrected using the off-peak intervals $E$ <772 eV, 788 eV < $E$ <789 eV and $E$ >800 eV.

**MOKE microscopy measurement**

The MOKE images were recorded using a custom-built wide-field MOKE

microscope. A background image was captured after applying a large negative or positive OOP magnetic field of 1 kOe. The background image was then subtracted from the subsequent images to obtain differential images with magnetic contrast. The velocity of the DW and skyrmion bubble was estimated by dividing the change in their position by the total duration of the current pulse. In the skyrmion bubble generation experiment, the racetrack device is initially magnetically saturated under a positive OOP magnetic field. Subsequently, a negative OOP magnetic field of approximately ~80 Oe is applied. While maintaining this reversed magnetic field, current pulses are introduced, which induce the nucleation and injection of skyrmion bubbles from the horn-shaped IP region. In the skyrmion bubble NOT gate experiment, skyrmion bubbles are first created using the same magnetic field and current pulse protocol as described above. Upon removal of the external magnetic field, we execute the skyrmion NOT operation.

The DW motion, DW inversion, skyrmion bubble generation and skyrmion logic were driven by current pulses generated with an AVTECH ultrahigh-speed pulse generator. The pulse generators can provide pulses of variable voltage and pulse width. The shape of the current pulse was detected by an oscilloscope (Agilent DSO9104A). The current densities were calculated by dividing the nominal voltage by the device resistance and cross-sectional area.

**Estimation of the speed of a logic operation**

We outline the procedure for estimating the speed of a logic operation in the NOT gates. Initially, we measure the DW velocity, $v_{DW}$, in the uniform OOP region of the racetracks. Subsequently, we determine the DW displacement, $L_{DW}$, across the NOT gate, following $N$ current pulses (Figures 2E and D). From these, we can obtain the time taken by the DW to transfer across the NOT gate, $t_{INV}$, and therefore the effective velocity of the DW, $v_{INV}$, as it transfers across the NOT gate and is inverted:

$$t_{INV} = N t_{pulse} - \frac{L_{DW} - L_{INV}}{v_{DW}}, \quad (1)$$

$$v_{INV} = \frac{L_{INV}}{t_{INV}}, \quad (2)$$

where $t_{pulse}$ is the duration of one current pulse and $L_{INV}$ is the length of the NOT gate. To obtain reliable values, we set $L_{INV}=1$ μm in the estimation in the experiment. With this method, we can determine $v_{DW}$ and $v_{INV}$ as a function of current density. The pulse length was decreased to 3 ns for high current densities to reduce heating.

**Electrical transport measurement**

The anomalous Hall effect measurements were performed in a standard four-probe geometry using a physical property measurement system (PPMS-9 T, Quantum Design).

**Micromagnetic simulation**

For the numerical study, we performed micromagnetic simulations to capture the features of a compensated ferrimagnet, comprising rate-earth and transition-metal elements. Simulations were carried out with the MuMax$^3$ codes using computation boxes with 1×1×1 nm$^3$ discretization, containing 512×1×1 cells for domain-wall motion and 384×128×1 cells for skyrmion devices, respectively. We assumed a simplified system that all the nearest neighbours are antiferromagnetically coupled with a negative exchange coupling constant $A_{ex}$ =-1.5×10$^{-11}$ J/m. More parameters are listed as following: saturation magnetization of two sublattices $M_{S,Co}$ =0.8 MA/m, $M_{S,Gd}$ =1.0 MA/m, uniaxial anisotropy energy $K_{u1}$ =600 kJ/m$^3$, $\theta_{SH}$ =0.1 and damping constant $\alpha$ =0.1. The interfacial DMI constant $D$ =0.4~1.4 mJ/m$^2$ for DW logic simulation and 2.0 mJ/m$^2$ for skyrmion logic simulation. For skyrmion logic simulation, values of $M_{S,Gd}$ =0.7, 0.8, 0.9 MA/m are selected to mimic different dominant elements of CoGd alloys.

**RESOURCE AVAILABILITY**

All data are available in the main text or the supplementary materials.

**Figures and figure captions**

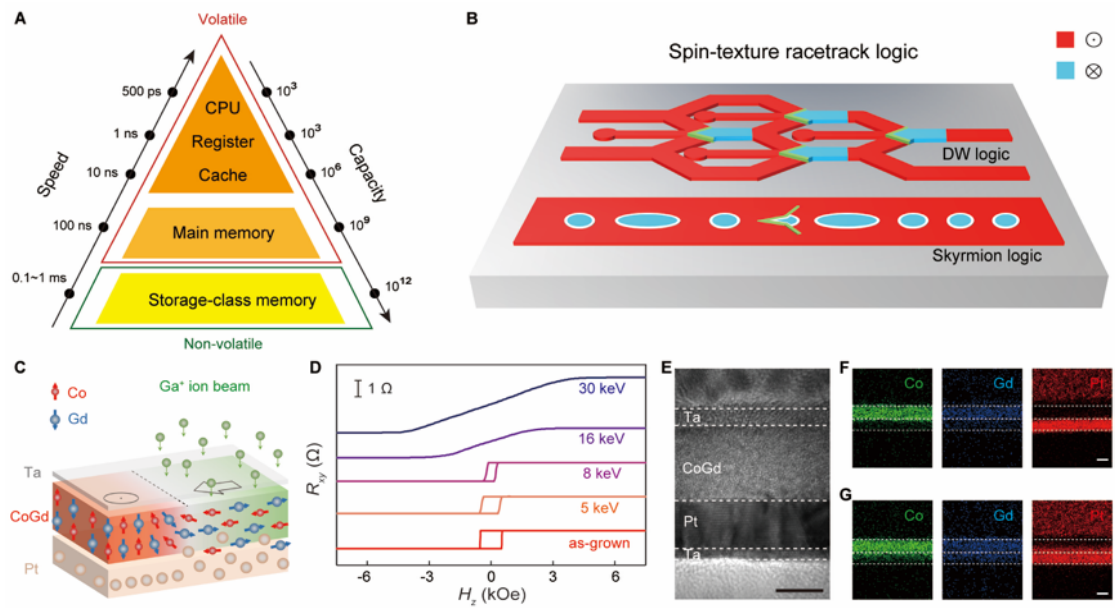

**Figure 1. Spin-texture racetrack logic and the principle for the modification of magnetic properties in the ferrimagnetic CoGd alloy.** (**A**) Schematic of the typical structure of the von Neumann computer hierarchy. (**B**) Schematic of spin-texture racetrack logic that encodes the information in nanosized spin textures of chiral DWs and skyrmion. (**C**) Schematic of the modification of magnetic anisotropy in a ferrimagnet by the Ga ion beam irradiation. (**D**) Anomalous Hall resistance ($R_{xy}$) of $Co_{78}Gd_{22}$ films after the irradiation with various ion energies and at the same irradiation dose of 57 µC/cm$^2$. (**E**) Cross-sectional TEM images of the as-grown $Co_{78}Gd_{22}$ film. As shown in the inset, the interfaces of the multilayer are indicated. (**F** and **G**) Co, Gd and Pt element mapping images of the $Co_{78}Gd_{22}$ film before (**F**) and after (**G**) the irradiation with the ion energy of 30 keV and at the irradiation dose of 57 µC/cm$^2$ by using energy dispersive X-ray spectroscopy (EDS). The interfaces are indicated by three white dashed lines from top to bottom respectively. All the scale bars are 5 nm.

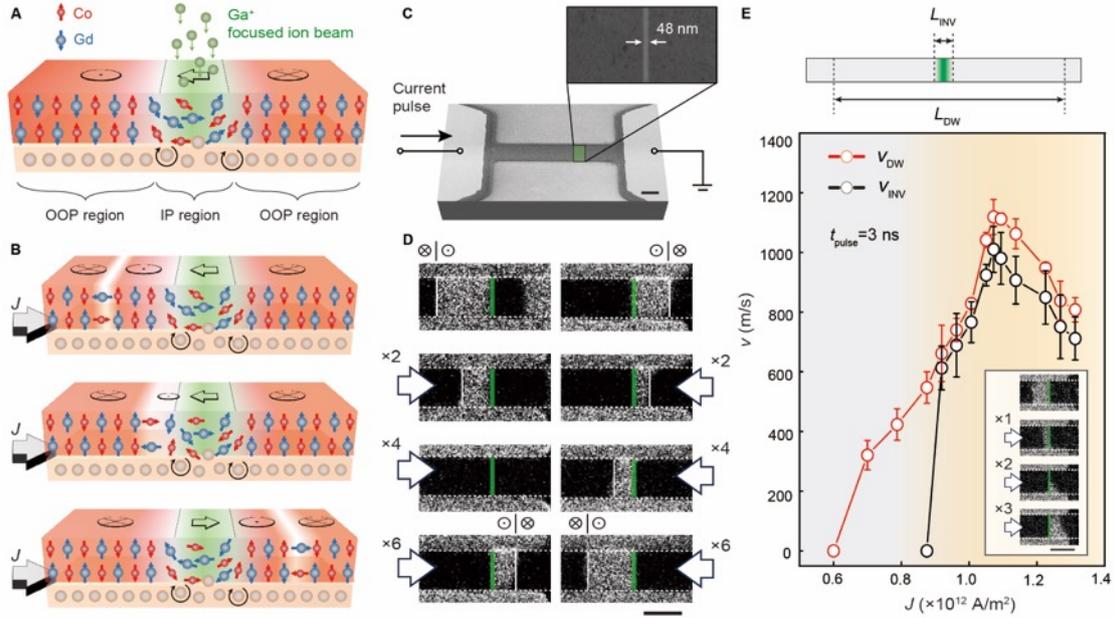

**Figure 2. Fast current-driven DW inversion in a chirally coupled ferrimagnet.** (**A**) Schematic of magnetic chiral coupling induced by the interfacial DMI in a ferrimagnet. After locally ion irradiation, the magnetizations of neighbouring OOP (un-irradiated; red-shaded) and IP (irradiated; green-shaded) regions align with a left-handed chirality in Pt/CoGd, forming a spin texture of ⊙←⊗. (**B**) Schematic illustrating the process of current-driven DW inversion. The white-shaded region is the DW, propagating along the direction of the electric current $J$. (**C**) SEM image of a DW inverter in a three-dimensional rendering of the DW measurement setup. As shown in the inset, the width of IP region in the DW inverter is indicated. (**D**) MOKE image sequence of DW inversion for a DW incident from the left (left column) and right (right column) side at a current density of $0.9\times10^{12}$ A/m$^2$. The edges of the magnetic racetracks are indicated by dashed white lines, the positions of DWs are indicated by white lines and the positions of the inverters are shown by green lines. The ×2, ×4, and ×6 indicate the number of current pulses. (**E**) DW velocity ($v_{DW}$) in a uniform OOP ferrimagnet and effective DW velocity ($v_{INV}$) in an inverter as a function of current density. The schematic to estimate $v_{INV}$ is shown in the inset. Error bars represent the standard deviation of the DW velocity for at least 5 repeated measurements. MOKE image sequence of DW inversion at the current density of $1.07\times10^{12}$ A/m$^2$ is shown in the inset. The bright and dark regions in the racetracks in the MOKE images correspond to ⊙ and ⊗ magnetization of the Co sublattice, respectively. The scale bars are 5 µm in the SEM images and 5 µm in the MOKE images.

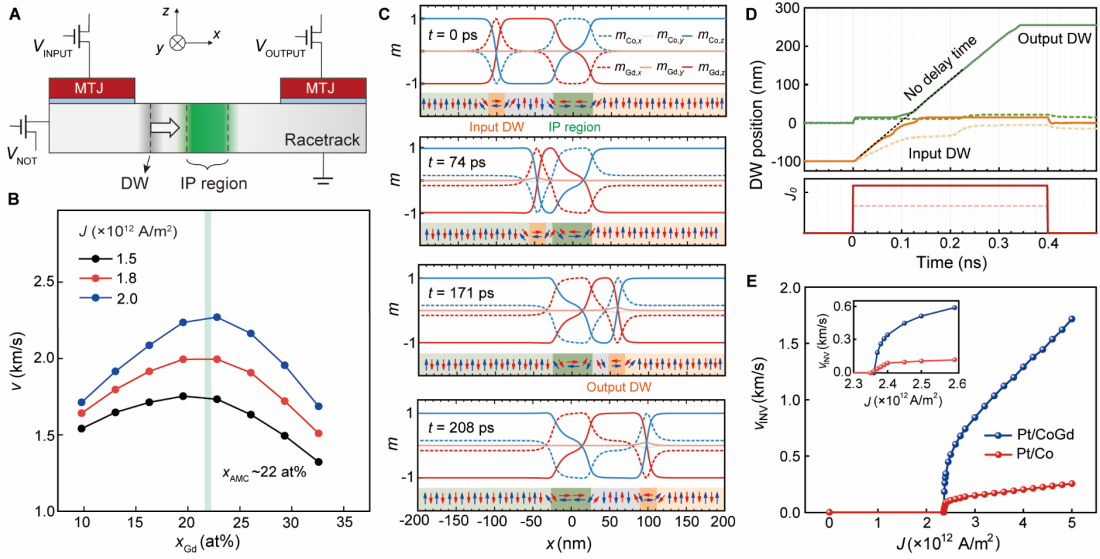

**Figure 3. Micromagnetic simulations for ultrafast DW logic.** (**A**) Schematic of spin-texture racetrack logic with two MTJs used to write/read the input/output. The gate voltages of $V_{INPUT}$, $V_{OUTPUT}$ and $V_{NOT}$ are used to write the input, read the output and execute the logic operation, respectively. (**B**) DW velocity as a function of the Gd atomic composition at various current densities of 1.5, 1.8 and 2.0×10$^{12}$ A/m$^2$. The atomic composition of angular momentum compensation state is indicated. (**C**) Snapshots of DW dynamics in the DW inverter at the current density of 3.5×10$^{12}$ A/m$^2$. The blue and red arrows indicate the direction of Co and Gd moments. (**D**) Time-resolved positions of the input and output DWs at the current densities below (dashed lines) and above (solid lines) the onset current density. The dashed black line indicates the nearly zero delay time between input and output DWs. (**E**) Effective DW velocity ($v_{INV}$) in the ferrimagnetic and ferromagnetic inverters as a function of the current density. The inset shows the DW velocities at low current density regime.

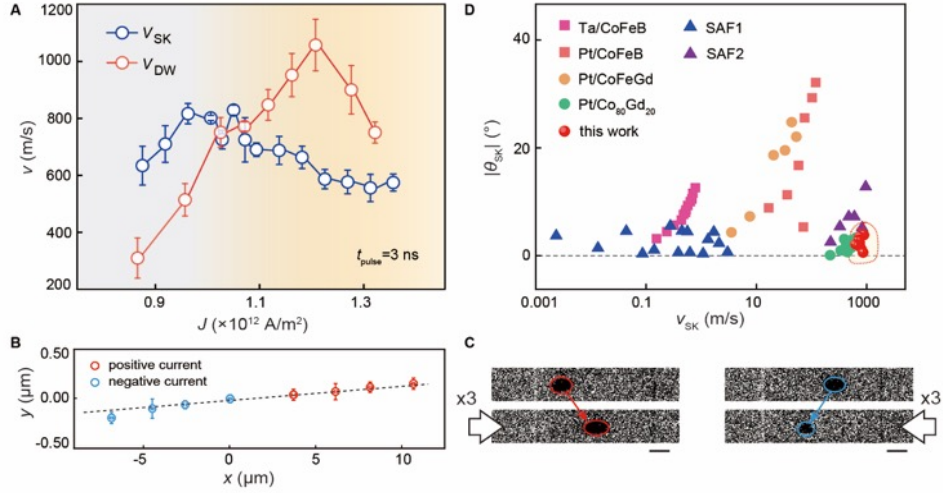

**Figure 4. Fast skyrmion motion with a small skyrmion Hall angle.** (**A**) $v_{SK}$ and $v_{DW}$ as a function of current density in the same ferrimagnetic racetrack. (**B**) Trajectory of skyrmion bubbles motion under the positive and negative current pulses along the x-axis with the current density of $1.01 \times 10^{12}$ A/m$^2$ and pulses width of 3 ns. Error bars represent the standard deviation of the centre of skyrmion bubble for at least 3 repeated measurements. (**C**) MOKE image sequence showing the current-induced displacement of skyrmion bubbles with positive (left) and negative (right) current pulses. (**D**) Skyrmion Hall angle $|\theta_{SK}|$ as a function of $v_{SK}$ for ferromagnetic (Ta/CoFeB[42], Pt/CoFeB[43]), synthetic antiferromagnetic (SAF1[50], SAF2[51]) and ferrimagnetic (Pt/CoFeGd[52], Pt/Co$_{80}$Gd$_{20}$[38]) skyrmion bubbles. The scale bars are 3 μm in the MOKE images.

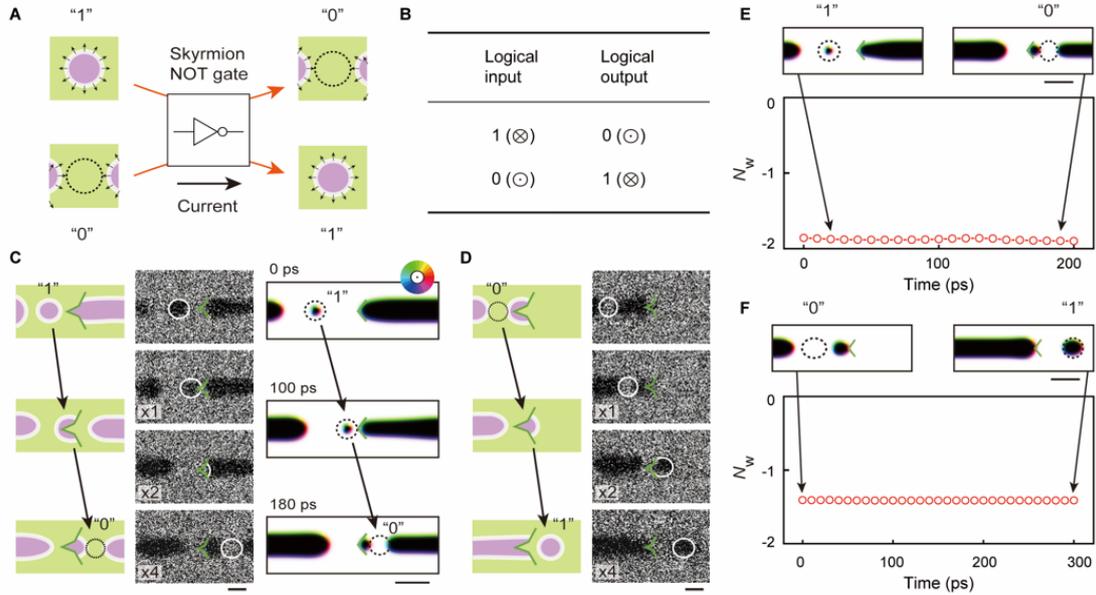

**Figure 5. Current-driven skyrmion logic.** (**A** and **B**) Schematics of a skyrmion NOT gate (**A**) and the corresponding truth table (**B**). (**C** and **D**) MOKE image sequence and corresponding schematics showing the skyrmion NOT operation for the logical input of "1" (**C**) and "0" (**D**) with the current density of $1.01 \times 10^{12}$ A/m$^2$ and pulses width of 3 ns. In the MOKE images, the boundaries of skyrmion bubbles are indicated by white dashed lines and the positions of the IP region are shown by green lines. The bright and dark regions in the MOKE images correspond to ⊙ and ⊗ magnetization of the Co sublattice, respectively. Micromagnetic simulation snapshots of the skyrmion NOT operation with the logical input of "1" at the current density of $1.2 \times 10^{12}$ A/m$^2$ are shown on the right of MOKE images in (**C**), with the direction of the magnetization indicated by the color wheel. The positions of the IP region are shown by green lines in the micromagnetic simulation snapshot. (**E** and **F**) Evolution of topological winding number $N_w$ during the NOT operation with the logical input of "1" (**E**) and "0" (**F**). The snapshots of micromagnetic simulations before and after the operation are indicated. The scale bars are 3 μm in the MOKE images and 40 nm in the simulation images.